\documentclass[twocolumn,amsmath,amssymb]{revtex4}
\usepackage[latin9]{inputenc}
\usepackage{amsmath}
\usepackage{amssymb}
\usepackage{graphicx}
\usepackage{esint}

\makeatletter
\@ifundefined{textcolor}{}
{%
 \definecolor{BLACK}{gray}{0}
 \definecolor{WHITE}{gray}{1}
 \definecolor{RED}{rgb}{1,0,0}
 \definecolor{GREEN}{rgb}{0,1,0}
 \definecolor{BLUE}{rgb}{0,0,1}
 \definecolor{CYAN}{cmyk}{1,0,0,0}
 \definecolor{MAGENTA}{cmyk}{0,1,0,0}
 \definecolor{YELLOW}{cmyk}{0,0,1,0}
}

\usepackage{mathrsfs}

\usepackage{amsfonts}
\usepackage{dcolumn}
\usepackage{bm}

\makeatother

\begin{document}

\title{Emergence of collective dynamical chirality for achiral active particles}

\author{Huijun Jiang, Huai Ding, Mingfeng Pu, Zhonghuai Hou}

\thanks{Corresponding Author: hzhlj@ustc.edu.cn}

\affiliation{Department of Chemical Physics \& Hefei National Laboratory for Physical
Sciences at Microscales, iChEM, University of Science and Technology
of China, Hefei, Anhui 230026, China}

\date{\today}

\begin{abstract}
Emergence of collective dynamical chirality (CDC) at mesoscopic scales
plays a key role in many formation processes of chiral structures
in nature, which may also provide possible routines for people to
fabricate complex chiral architectures. So far, most of reported CDCs
are found in systems of active objects with individual structure chirality
or/and dynamical chirality, and whether CDC can arise from simple
and achiral units is still an attractive mystery. Here, we report
a spontaneous formation of CDC in a system of both dynamically and
structurally achiral particles motivated by active motion of cells adhered on a substrate.
Active moving, confinement and hydrodynamic interaction are found
to be the three key factors. Detailed analysis shows that the system
can support abundant collective dynamical behaviors, including rotating
droplet, rotating bubble, CDC oscillation, array of collective rotation,
as well as interesting transitions such as chirality transition, structure
transition and state reentrance.
\end{abstract}

\maketitle
\section{Introduction}
Collective motion of natural or artificial
micro(meso)scopic active objects has attracted growing research interests
due to its ubiquity and importance in many systems\cite{marchetti2013hydrodynamics,elgeti2015physics,wioland2016ferromagnetic,lushi2014fluid,goto2015purely,szabo2006phase}.
One example is that groups of cells or their fragments can undergo
active motion\cite{szabo2006phase,ananthakrishnan2007forces,weber2012mechanoresponsive,sato2015cell,tambe2011collective},
and a variety of fundamental processes in development, health and
disease depend on such coordinated motions\cite{levental2009matrix,bianco2007two,friedl2009collective,montell2008morphogenetic}.
Since active systems can be driven far from equilibrium by continuously
consuming energy supplied internally or externally, they are capable
of completely altering the collective dynamical behaviors of interacting
motile particles in a fashion that is forbidden for non-active particles.

Collective dynamical chirality (CDC) --mirror asymmetry of the collective
motion of motile objects-- is one of such interesting behaviors supported
in active systems. On the one hand, CDC plays a crucial role in
many processes in nature. For instance, establishment of left-right
asymmetry in embryonic development, one of the most intriguing biological
phenomena, involves coordinated activity of many cells\cite{blum2014evolution,coutelis2014diversity}
where the ability of cells to distinguish between left and right is
evident in systems of chiral patterns formed by collective motion
of identical cells confined in circular island or ring/stripe-shaped
micropatterns\cite{wan2011micropatterned,yamanaka2015rotating,tamada2010autonomous,xu2007polarity}.
On the other hand, CDC may also inspire new routines for fabrication
of complex chiral architectures by dynamically self-assembling simple
and achiral building blocks\cite{guerrero2011individual}, e.g., chiral
clusters of asymmetric colloidal dimers have been successfully assembled
by using alternating current electric fields\cite{ma2015electric}.
Revealing how CDC arises from groups of active units is then very
important for the understanding of the formation mechanism.

So far, CDC can be found in systems of active objects with individual
structure chirality and/or individual dynamical chirality. For example,
CDC has been reported in several experiments where vortexes were observed
for microtubules\cite{sumino2012large}, actin flaments\cite{schaller2010polar,woodhouse2012spontaneous},
or sperm cells\cite{riedel2005self} moving on a planar surface. It
is believed that individual dynamical chirality might be caused by
the rotation of the microtubule around its axis\cite{kagami1992translocation}
or by the special slender shape\cite{jiang2014motion}, while interactions
between active objects help to align their moving direction\cite{sumino2012large,riedel2005self,jiang2014hydrodynamic}.
Besides, man-made catalytical nanorods\cite{dhar2006autonomously,qin2007rational},
self-motile colloids\cite{gibbs2011geometrically} or rotating disks\cite{goto2015purely}
can also be dynamically chiral in the form of swimming in circles,
and hydrodynamic interaction can synchronize them to form CDC\cite{goto2015purely}. There is also a recent study of elliptical active particles where particles can rotate in a circular confinement\cite{lushi2014fluid}. For
structurally chiral objects, dynamical chirality will arise when they
are driven by external fields through potential landscapes, and CDC
can provide an efficient method for chirality sorting\cite{kostur2006chiral,speer2010exploiting,bogunovic2012chiral,nourhani2015guiding}.
Very recently, a metastable CDC is reported in a system with achiral interaction\cite{Breier2016spontaneous}. Since such a metastable chiral state will relax to a more stable state for a finite temperature, it is still a very attractive mystery that whether stable CDC can emerge in systems of simple particles without both individual structure
chirality and dynamical chirality.

In this paper, we employ a model motivated by active motion of cells adhered on a surface
in fluid environment to address such a question. The model consists of three elementary ingredients,
i.e., achiral active moving, confined space for particle motion and
hydrodynamic interaction (HI) between particles, to avoid other complexity
such as special shape or structure chirality in real systems which
perplexes us to understand the fundamental mechanism underlying formation
of CDC\cite{guerrero2011individual}. Remarkably, we find that CDC
emerges spontaneously in the form of collective rotating for active
forces larger than a critical value, near which an interesting oscillation
between clockwise and anti-clockwise rotation is observed. Detailed
analysis reveals that confinement and HI, along
with the active motion of particles, are sufficient for the formation
of CDC, while other details such as confinement shape and boundary
condition are not relevant. Moreover, phase diagram shows that the
system supports abundant collective states, e.g., two distinct states
of CDC, a rotating droplet state and a rotating bubble state, are
identified by a structure transition of CDC states, and interesting
transition behaviors such as state reentrance from fluid-like state
to rotating droplet then back to fluid-like state can also be observed.
In addition, we find that the number of collective rotation is determined
by the width/ length ratio of the confinement: A single droplet can
be found in squares, while arrays of multiple ones are observed in
rectangles.

\section{Model}
We consider a system consisting of $N$ active spherical
particles moving in a two-dimensional rectangle confined space of
size $L\times W$, where $L$ and $W$ are the length and width, respectively.
Active motion of particles is realized by exerting a constant force
$f_{0}$ along the internal active directions, which mimics cell movement adhered
on a surface\cite{ananthakrishnan2007forces}. The only direct
interaction between particles is exclusive volume effect taken into
account by a Weeks-Chandler-Andersen potential $U(\boldsymbol{\mathbf{r}}_{ij})=4\epsilon\left\{ \left(2a/r_{ij}\right)^{12}-\left(2a/r_{ij}\right)^{6}\right\} +\epsilon$
existing only if $r_{ij}<2\sqrt[6]{2}a$, where $a$ is the effective
repulsion radius, $\boldsymbol{r}_{i}$ the position of particle $i$,
and $r_{ij}$ the distance between $i$-th and $j$-th particle. Particles
can also interact with each other indirectly by long-range HI through
the ambient fluid, where the force on $i$-th particle generated by
the fluid is $\mathbf{F}_{i,fl}=-\gamma[\boldsymbol{\dot{r}}_{i}-\mathbf{u}(\boldsymbol{r}_{i},t)]$
with $\gamma$ the friction coefficient and $\mathbf{u}(\boldsymbol{r}_{i},t)$
the fluid velocity at location $\boldsymbol{r}_{i}$. In our work,
$\mathbf{u}(\boldsymbol{r}_{i},t)$ is calculated by a stochastic
lattice Boltzmann method where particles are treated as point-like
ones\cite{ahlrichs1998lattice}. Simulation details of the lattice Boltzmann method can be found in the supplemental information$^\dag$.The equations for translational motion
of particles with mass $m$ are then

\begin{equation}
m\mathbf{\ddot{r}}_{i}=f_{0}\mathbf{n}_{i}+\mathbf{F}_{i,fl}-\sum_{j=1}^{N}\frac{\partial U(\mathbf{r}_{ij})}{\partial\mathbf{r}_{ij}}+\mathbf{\xi}_{i},\:i=1,...,N.\label{eq:translation}
\end{equation}
Herein, $\mathbf{\xi}_{i}(t)$ denotes the fluctuation force satisfying
the fluctuation-dissipation relation $<\mathbf{\xi}_{i}(t)\mathbf{\xi}_{j}(t')>=2\gamma k_{B}T\delta(t-t')\delta_{ij}$
where $k_{B}$ is the Boltzmann constant, $T$ denotes the temperature,
and $\boldsymbol{n}_{i}=(\cos\theta_{i},\sin\theta_{i})$ is the direction
of active force. Besides, the angle $\theta_{i}$ is steered towards
the direction of total force on $i$-th particle by the rule similar
to the one in Ref\citenum{szabo2006phase,mones2015anomalous}:

\begin{equation}
\dot{\theta}_{i}=(1/\tau)[\arg(\ddot{\mathbf{r}}_{i})-\theta_{i}]+\mathbf{\zeta}_{i}\label{eq:rotation}
\end{equation}
where $\arg(\mathbf{\ddot{r}}_{i})$ is the angle of total force vector,
and $\mathbf{\zeta}_{i}(t)$ is the rotational fluctuation satisfying$< \mathbf{\zeta}_{i}(t)\mathbf{\zeta}_{j}(t')>=3k_{B}T\delta(t-t')\delta_{ij}/(2a^{2}\gamma)$.
Such a steering rule is motivated by the fact that cells can respond
to mechanical forces\cite{kozlov2007model,yam2007actin,tambe2011collective},
which is also consistent with reported positive feedback regulation
of front-rear cell polarity by actual cell displacements\cite{dawes2007phosphoinositides,szabo2010collective}.
The resistance time $\tau$ for active orientation measures the ability
of particles resisting such external steering. For a small $\tau$, active particles
will yield to the steering very fast, while for a large enough $\tau$ they tend
to keep their internal random behavior similar
to the conventional ones without steering rule\cite{speck2014effective,ni2015tunable}. We rescale the density, time and
length by the particle density, the simulation time step of the lattice
Boltzmann method, and the grid length of the lattice, respectively.
We fix $k_{B}T=10^{-7}$, $\epsilon=5\times10^{-4}$, $a=0.75$, and
$\gamma=(32/3)a\nu\varrho$ with fluid viscosity $\nu=0.1$ and density
$\varrho=1$, for which the system will reach a fluid-like state in
the absence of active force. The resistance time, active force, number
of particles and size of the space are $\tau=1$, $f_{0}=1.25\times10^{-5}$,
$N=2500$ and $L=W=100$ (corresponding to a volume fraction about
$0.442$), if not otherwise stated. The interaction between active
particles and confined boundary is realized by bounce-back rule.

\section{Result}
\begin{figure}
\begin{centering}
\includegraphics[width=1.0\columnwidth]{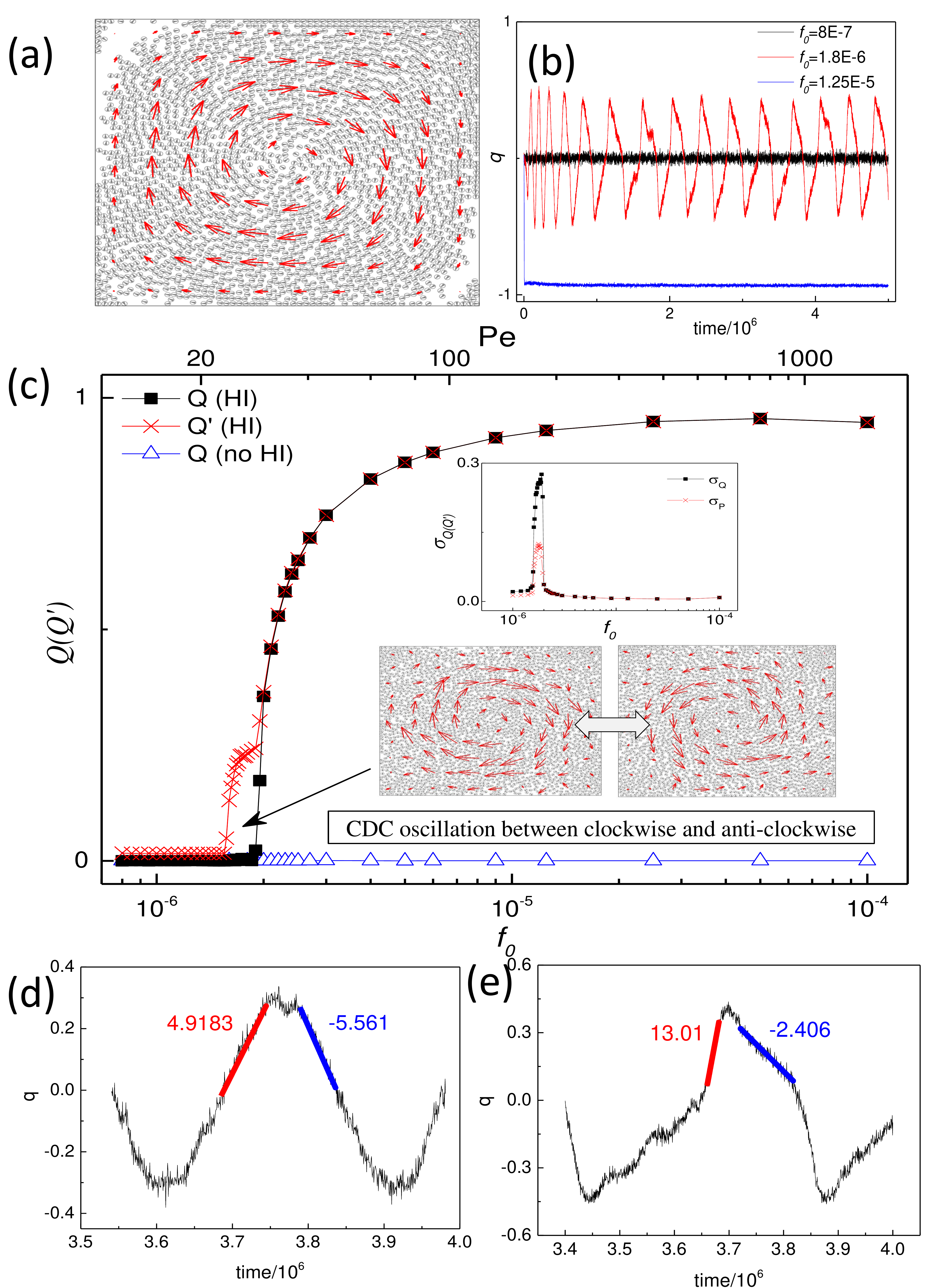}
\par\end{centering}

\caption{Spontaneous emergence of collective dynamical chirality. (a) Typical snapshot
of CDC state for $f_{0}=1.25\times10^{-5}$. Red arrows are locally averaged velocities of particles in the nearest $10\times10$ grids (normalized by the maximal
one). (b) Time series of $q$ for $f_{0}=8\times10^{-7}$, $1.8\times10^{-6}$ and $1.25\times10^{-5}$. (c) Dependence of order parameters $Q$ and $Q'$
on the active force $f_{0}$. Standard deviation of $Q$ is plotted in the top inset and typical snapshots of clockwise and anti-clockwise rotation during chirality oscillation in the bottom. For comparison, $Q$ without hydrodynamic interaction and its standard deviation are also plotted. For (d) $f_0=1.67\times10^{-6}$ and (e) $f_0=1.8\times10^{-6}$, time scales for onset of CDC away from $q = 0$ and decay towards $q = 0$ are shown by slopes of red and blue lines, respectively.}
\label{fig:chirality}
\end{figure}

To begin, we investigate how the collective motion
of particles depends on the magnitude $f_{0}$ of the active force.
For a small active force, e.g., $f_{0}=8\times10^{-7}$, the system
is still fluid-like where particles move randomly. Quite interestingly,
a CDC state emerges spontaneously if the active force becomes large
enough. A typical snapshot of such a state for $f_{0}=1.25\times10^{-5}$
is presented in Fig.\ref{fig:chirality}(a) where all the particles
rotate collectively around the center of the space $\mathbf{r}_{c}=(L/2,W/2)$.
To quantitatively characterize CDC of the system, we define an order
parameter

\begin{equation}
q=\frac{1}{N}\sum_{i=1}^{N}\varphi_{i}.
\end{equation}
Here, $\varphi_{i}=\omega_{i}/|\omega_{i}|$ denotes the ``dynamical-chirality
spin'' of particle $i$, where $\omega_{i}=(\mathbf{r}_{i}-\mathbf{r}_{c})\times\dot{\mathbf{r}}_{i}/|(\mathbf{r}_{i}-\mathbf{r}_{c})|^{2}$
is the angular velocity of $i$-th particle relative to $\mathbf{r}_{c}$.
Note that $\varphi_{i}$ equals to $1$ for anti-clockwise rotation
and $-1$ for clockwise one. Time-dependencies of $q$ for $f_{0}=8\times10^{-7}$
and $1.25\times10^{-5}$ are plotted in Fig.\ref{fig:chirality}(b).
It can be observed that $q$ fluctuates around a fixed value after
a quick relaxation, indicating that the system can finally reach a
stable steady state. Time-averaged $q$
equals $0$ for the fluid-like state, and is of a negative (or positive) value for the CDC
state with collective clockwise (or anti-clockwise) rotation.

To obtain a global picture for how CDC emerges as $f_{0}$ increases, magnitude
of time-averaged $q$,
\begin{equation}
Q=\left|\underset{_{t_{0}\rightarrow\infty}}{\lim}\frac{1}{t_{0}}\int_{0}^{t_{0}}q(t)dt\right|
\end{equation}
as a function of $f_{0}$ is drawn in Fig.\ref{fig:chirality}(c). For spherical active particles, there is also another parameter to measure the activity of particles, i.e., the P\'{e}clet number $Pe=|f_0|a/(k_BT)$. To be comparison, the corresponding value of $Pe$ are also shown in the top axis in Fig.\ref{fig:chirality}(c). Clearly, a continuous-like transition from fluid-like state to CDC
state induced by particle activity is observed: $Q$ is nearly zero
for small forces and quickly increases to be nearly $1$ for $f_{0}$
larger than a threshold $f_{c}\simeq 1.9\times10^{-6}$. By defining the standard
deviation as $\sigma_{Q}=\sqrt{(1/t_0)\int^{t_0}_0 [q(t)-\bar{q}]^2 dt}$ where $\bar{q}$ is the mean value of $q(t)$, $\sigma_Q$ exhibits a clear-cut peak as shown
in the top inset of Fig.\ref{fig:chirality}(c). The presence of $f_c$
may be understood by the following observations. By taking a close look at the top-right and bottom-left corner in Fig.\ref{fig:chirality}(a), it can be seen that collective rotation can lead
to accumulation of particles near the boundary at those corners. The accumulation will in return provide an obstacle for particles to rotate collectively and consequently an effective barrier for emergence of CDC.

Remarkably, we also observe an interesting oscillation of CDC. As
depicted in Fig.\ref{fig:chirality}(b), particles rotate periodically
between clockwise and anti-clockwise for an active force slightly
smaller than $f_{c}$, e.g., $f_{0}=1.8\times10^{-6}$, whose typical
snapshots are shown in the bottom-left and bottom-right insets of
Fig.\ref{fig:chirality}(c). The formation of CDC oscillation may be due to
competition of the onset process of CDC away from $q = 0$ and decay process towards $q = 0$. In Fig.\ref{fig:chirality}(d),
rates of these two processes are presented by absolute values of the time-series slope for $f_0=1.67\times10^{-6}$, where the onset rate is about $4.92$ while the decay one is $5.56$. When the active force $f_0$ approaches the threshold $f_c$, for example $f_0=1.8\times10^{-6}$, the onset process is accelerated to be of a slope $13.1$, in the meanwhile, the decay one decreases to $2.406$ (Fig.\ref{fig:chirality}(e)). The observation implies that CDC is hard to onset and easy to decay for small active forces, and the decay rate may approach 0 and only emergence of CDC can be observed for large enough active forces. Thus, an appropriate active force can lead to CDC oscillation. To elucidate more clearly the detailed mechanism for the formation of CDC oscillation, a follow-up study may needed.

To identify the region of CDC oscillation, the time-averaged
magnitude of $q$, $Q'=\lim_{t_{0}\rightarrow\infty}(1/t_{0})\int_{0}^{t_{0}}|q(t)|dt$,
is presented in Fig.\ref{fig:chirality}(c). It can be found that
$Q'$ is overlapped with $Q$ very well for small or large active
forces. In the CDC region where active force is in the range $1.55\times10^{-6}<f_{0}<f_{c}$,
$Q=0$ indicates that there is no time-averaged CDC, while $Q'$ is
larger than zero obviously, demonstrating that particles do rotate
collectively for a given time. The standard deviation of $Q'$, $\sigma_{Q'}=\sqrt{(1/t_0)\int^{t_0}_0 [|q(t)|-\bar{|q|}]^2 dt}$ where $\bar{|q|}$ is the mean value of $|q(t)|$, is
also plotted in the top inset of Fig.\ref{fig:chirality}(c), which
shows a peak at $f_{0} \simeq 1.9\times10^{-6}$ as same as
the one of $Q$, demonstrating that chirality oscillation is not a new
dynamical phase of the system.

One may be wondering what are the key ingredients that lead to the above interesting observations.
In fact, we find that besides the active driving, the long rang HI and space confinement are
two other necessary conditions for emergence of CDC as well as the CDC-oscillation.
To show this, we have performed parallel simulations with the same parameter settings as above
but with the HI turned off by using Brownian dynamics where diffusion
coefficient of a free-diffusion particle is ensured to be the same.
The obtained $Q$ without HI is plotted in Fig.\ref{fig:chirality}(c)
to be compared with the one with HI. Clearly, there is no CDC can be observed
for all range of parameters. We also repeat similar simulations for collective
motion of particles without confinement, and no CDC is found, too.
What's more, other rules of interaction between active particles and
confined boundary such as reflecting rule are also tested, and our
findings are not sensitive to the boundary condition. In short, HI
and confinement, along with activity of particles are the three key
factors for the emergence of CDC. Notice that, active particles we used here are force monopoles to mimic cell movement adhered on a surface. For microswimmers suspended in fluid, they form at least force dipoles. We have repeated similar simulations for microswimmers, and no CDC was observed for the parameters presented here.

\begin{figure}
\begin{centering}
\includegraphics[width=1\columnwidth]{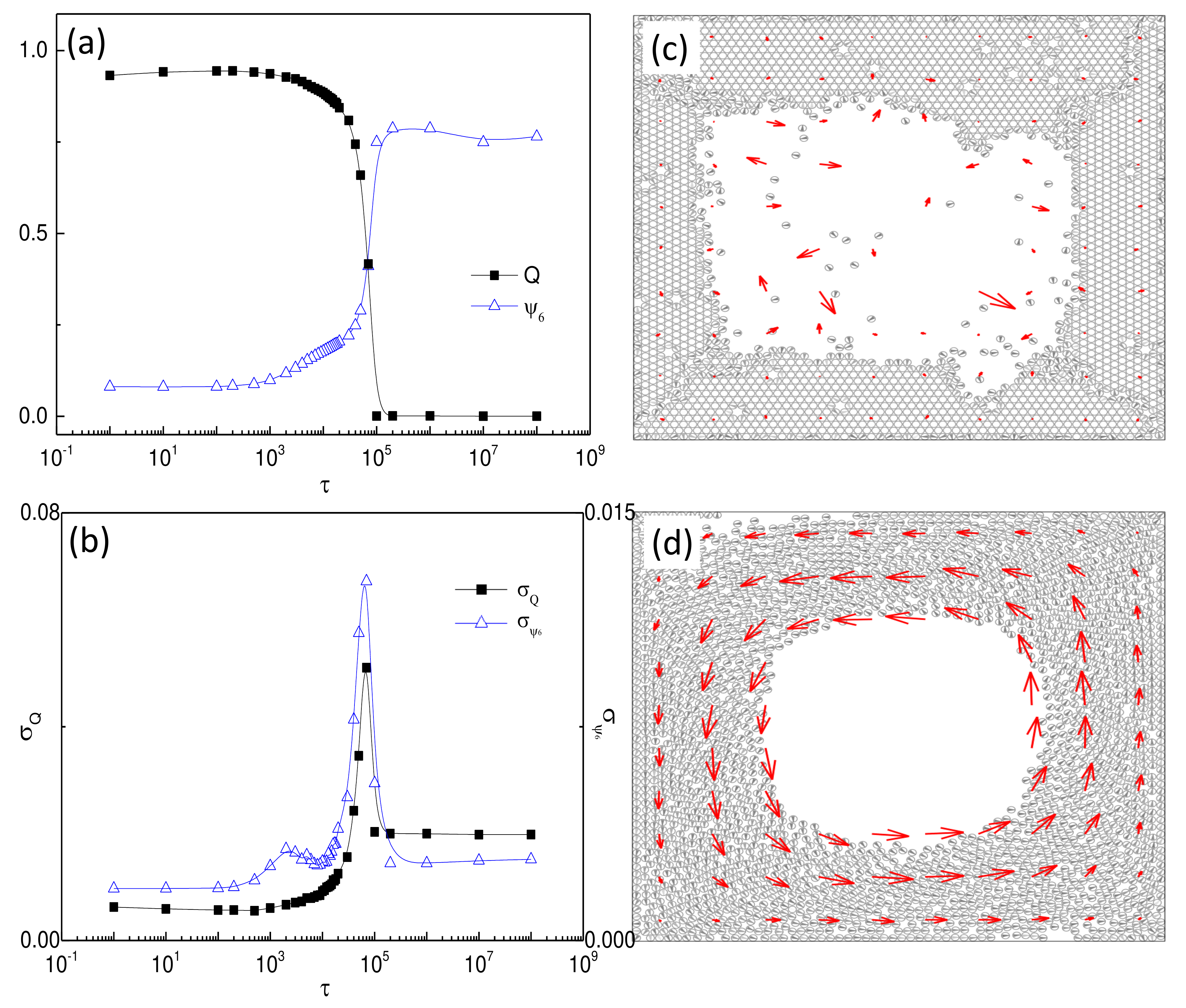}
\par\end{centering}

\caption{(a) Dependence of order parameters $Q$ and $\Psi_{6}$ on resistance
time $\tau$. (b) Standard deviation of order parameters as functions of $\tau$.
(c) Typical snapshots of crystal-like state for $\tau=10^{8}$ and (d) rotating bubble state
for $\tau=10^{4}$. The locally averaged velocities
of particles normalized by the maximal one are presented by red arrows. }
\label{fig:tau}
\end{figure}

It is noted that, dynamics of active particles may change dramatically
for different resistance time for active direction, thus, we now
try to figure out how $\tau$ affects collective motion of active
particles. In Fig.\ref{fig:tau}(a), $Q$ as a function of $\tau$
is plotted by fixing $f_{0}=1.25\times10^{-5}$, where a transition
between CDC and an achiral state can also be found. A typical snapshot
of the achiral state is presented in Fig.\ref{fig:tau}(c). Different
to the fluid-like state, the achiral state observed here is crystal-like
where particles are arranged in hexagonal ordering. The ordering can
be measured by\cite{stenhammar2015activity}

\begin{equation}
\Psi_{6}=\frac{1}{N}\sum_{m=1}^{N}\frac{1}{N_{m}}\sum_{l=1}^{N_{m}}\exp(6i\theta_{ml}),
\end{equation}
where $i$ is the imaginary unit, $\theta_{ml}$ is the angle between
an arbitrary reference axis and the displacement vector between particles
$m$ and $l$, and the sum runs over the nearest $N_{m}=6$ particles
within a cutoff radius of $2.6a$ from particle $m$ (for particles
adjacent to the confined boundary, only $N_{m}=4$ neighbors are needed
to form hexagonal ordering). In Fig.\ref{fig:tau}(a), $\Psi_{6}$
increases from a value near $0$ to about $1$ as $\tau$ increases,
indicating a structure transition when collective motion of particles
changes CDC state to crystal-like state.

If one takes a closer look at the dependence of $\Psi_{6}$ on $\tau$,
it can be found that there seems to be a shoulder before the transition
happens. For more detailed information, standard deviations $\sigma$
of $Q$ and $\Psi_{6}$ are given in Fig.\ref{fig:tau}(b). As expected,
peaks are observed for both $\sigma{}_{Q}$ and $\sigma_{\Psi_{6}}$
at $\tau_{c}\simeq7\times10^{4}$, corresponding to the transition
of both chirality and structure from CDC state to crystal-like state.
It is quite interesting that there is also another peak of $\sigma_{\Psi_{6}}$
at $\tau_{s}\simeq 2\times10^{3}$ where no peaks of $\sigma_{Q}$
are found, i.e., the CDC state undergoes a structure transition without
loss of chirality. By comparison between the snapshot for $\tau=1<\tau_{s}$
in Fig.\ref{fig:chirality}(b) and a typical snapshot presented in
Fig.\ref{fig:tau}(d) for $\tau_{s}<\tau=10^{4}<\tau_{c}$, we can
mark these two states of CDC as rotating droplet for the former and
rotating bubble for the later.

\begin{figure}
\begin{centering}
\includegraphics[width=0.8\columnwidth]{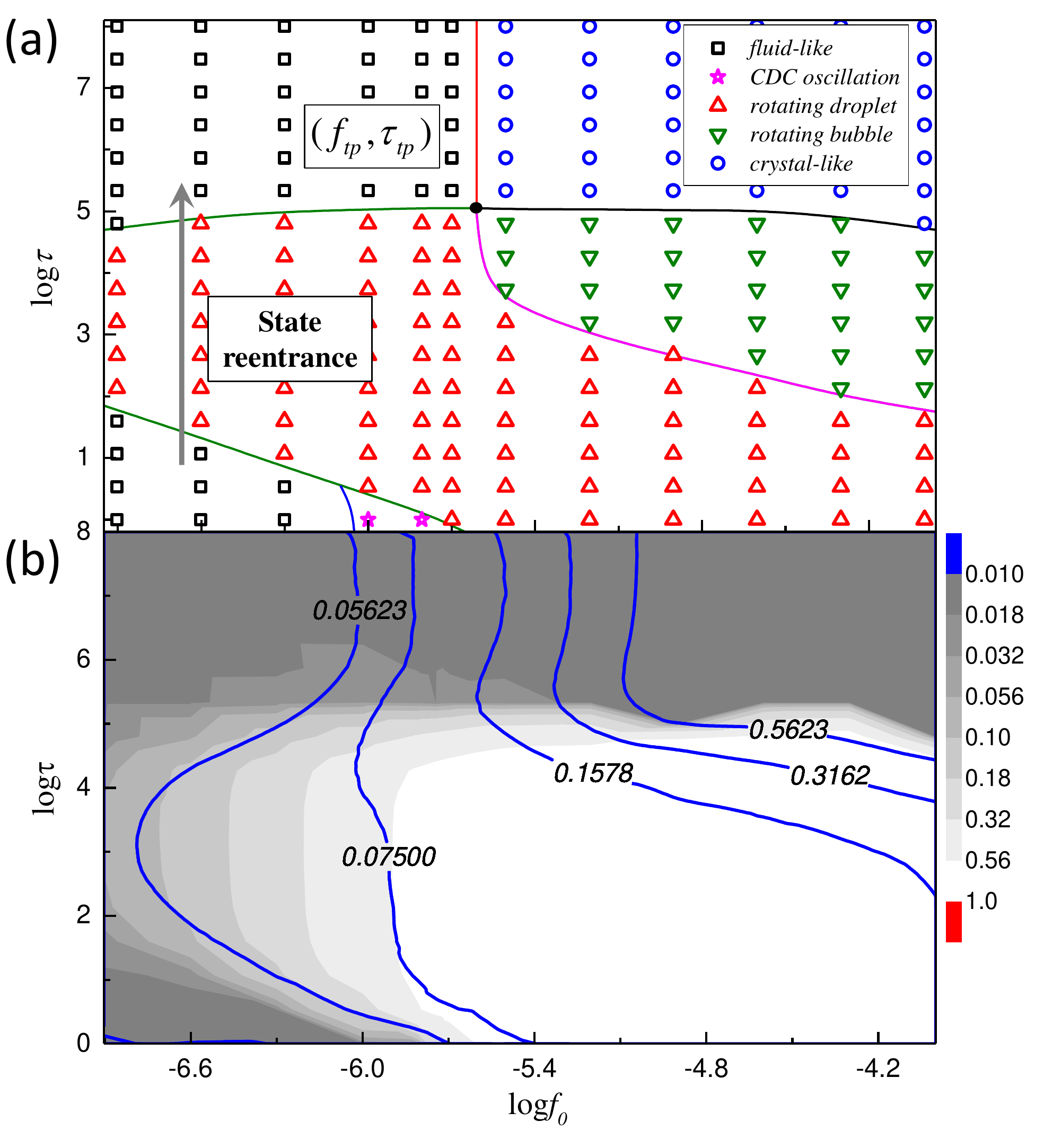}
\par\end{centering}

\caption{(a) Phase diagram on $f_{0}-\tau$ plane. There is a triple-point-like
point $(f_{tp},\tau_{tp})$ above which no CDC can be formed for $\tau>\tau_{tp}$,
and a state reentrance from fluid-like state to rotating droplet state
then back to fluid-like state can be found as indicated by the gray
arrow. (b) The color plot for $Q$ and the contour plot for $\Psi_6$ in the same parameter region as in (a). The color from gray to white indicates the value of $Q$ from $0.01$ to $1$ and lines with labels present corresponding values of $\Psi_6$. }
\label{fig:PD}
\end{figure}

In order to explore fully how parameters affect particles' collective motion, a phase diagram in $f_{0}-\tau$ plane is obtained by extensive simulations (Fig.\ref{fig:PD}(a)). Several interesting remarks can be made. Firstly, there is a triple-point-like
point located at $(f_{tp},\tau_{tp})$ where CDC state meets both fluid-like and crystal-like state. For $\tau<\tau_{tp}$,
CDC can arise spontaneously, while for larger $\tau$s only fluid-like state and crystal-like state can be observed.
Notice that the steering rule in Eq.(\ref{eq:rotation}) can be neglected for large enough $\tau$, the observation
is in good agreement with findings reported in literature\cite{speck2014effective,ni2015tunable}. Secondly, structure transition between the two states of CDC, rotating droplet and rotating bubble, occurs only for $f_{0}>f_{tp}$, below which rotating droplet is the
sole CDC state. Lastly, the system can also support other interesting state transition behaviors. As indicated by the gray arrow in Fig.\ref{fig:PD}, a state reentrance can be found as $\tau$ increases for $f_{0}<f_{tp}$, i.e., particles are firstly fluid-like for small $\tau$s then change to be of CDC for intermediate $\tau s$, then back to be fluid-like when $\tau$ is large enough. The corresponding color plot for $Q$ and contour plot for $\Psi_6$ are also presented in Fig.\ref{fig:PD}(b). It can be found that, as $\tau$ passes by $\tau_{tp}$, transition from the fluid/crystalline to the chiral state results in a rapid increase in $Q$ for large $f_0$ as already shown in Fig.\ref{fig:tau}. Moreover for small $f_0$ the change is much more subdued, indicating that the two fluid-like areas are connected for very small $f_0$. When $f_0$ crossing $f_{tp}$, i.e. the system goes from the fluid to the crystalline state, we see that the $\Psi_6$ increases rapidly, showing a crystallisation/melting transition due to activity. Furthermore, the rotating bubble has a higher $\Psi_6$ than the rotating droplet, which is probably due to the walls facilitating a hexagonal order.

\begin{figure}
\begin{centering}
\includegraphics[width=0.8\columnwidth]{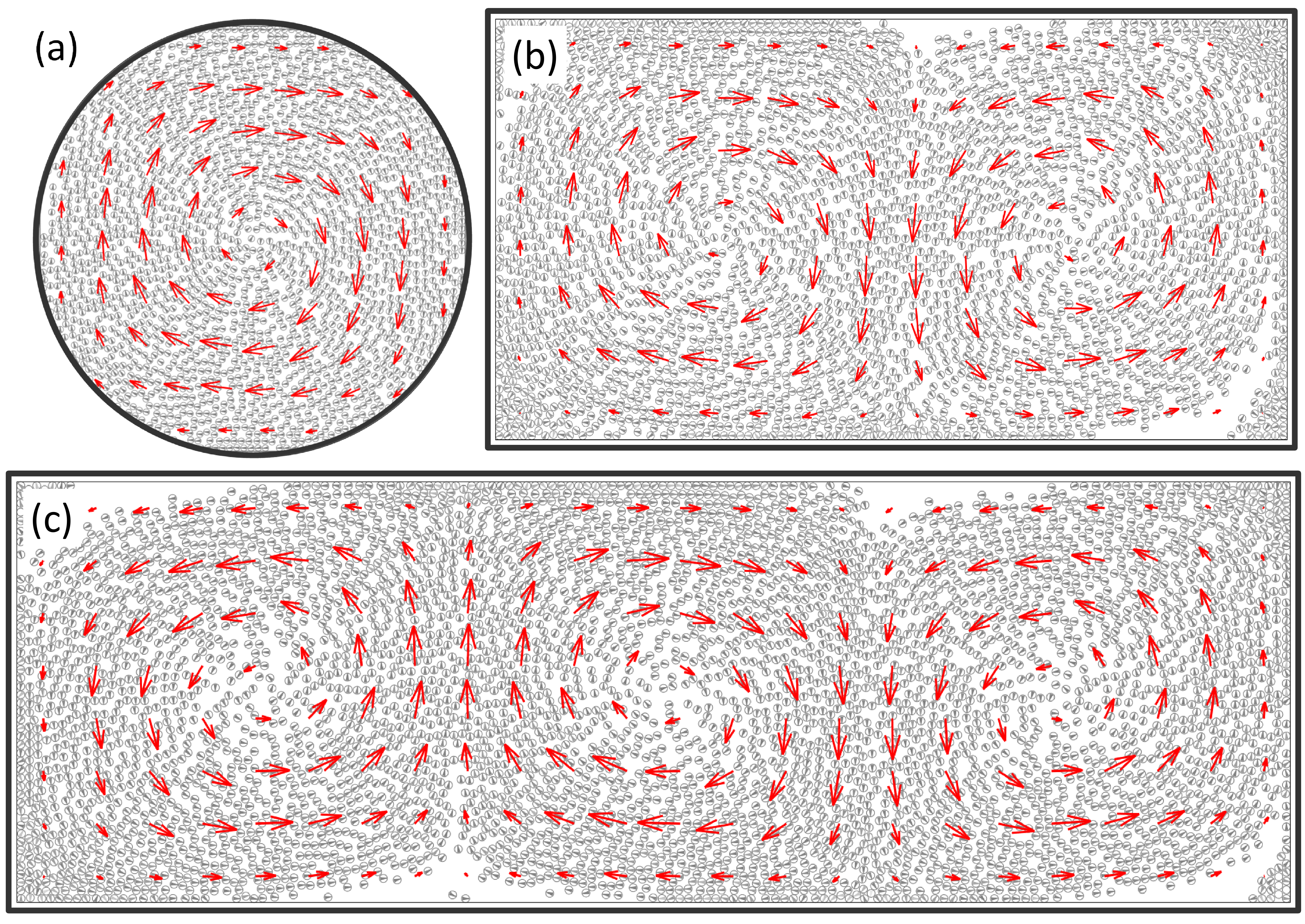}
\par\end{centering}
\caption{Collective dynamical chirality for different boundary shapes. (a)
Circle with diameter $100$, and rectangles with (b) $160\times80$
and (c) $240\times80$ for $f_{0}=1.25\times10^{-5}$ and $\tau=1$.}
\label{fig:BC}
\end{figure}

At last, effects of confinement shape on CDC are also considered. A
typical snapshot for collective motion of particles in a circular confined
space with diameter $100$ is shown in Fig.\ref{fig:BC}(a). Similarly,
CDC emerges for the same parameters as in the square one, indicating
that formation of CDC is not sensitive to the confinement shape. Dynamics
of particles in confinement with different length/width ratio is also
investigated. Typical snapshots for $L\times W=160\times80$ and $240\times80$
are presented in Fig.\ref{fig:BC}(b) and Fig.\ref{fig:BC}(c), respectively,
where number of particles is set to be $N=3200$ for the former and
$N=4800$ for the later to keep the volume fraction unchanged. Interestingly,
array of vortexes with opposite chirality for adjacent ones is observed,
and the number of vortexes seems to be proportional to the length/width
ratio.

\section{Conclusion}
In summary, it was revealed that active motion, space
confinement and hydrodynamic interaction are sufficient for the emergence of CDC
in a system of active particles without individual structure chirality and dynamical chirality.
CDC states were found to be formed via a chirality transition from other achiral state such as fluid-like state
or crystal-like state, while they can also undergo a structure transition to form two distinct states, i.e.,
a rotating droplet and a rotating bubble. Phase diagram showed that CDC formation is controlled by the active force and
the resistance time for particles to maintain their internal motion. More interestingly, CDC oscillation can also be supported by the system. Formation of CDC oscillation may be due to a competition between the onset process of CDC away from $q = 0$ and decay process towards $q = 0$. Since emergence of CDC underlies many formation processes of
chiral structures, our finding may inspire experimental studies to explore new routines for fabrication of complex chiral architectures
by simple and achiral units, and shed light on the understanding of chirality formation in other complex systems such as establishment of left-right
asymmetry in embryonic development.

\section{Acknowledgments}
This work is supported by National Basic Research Program of China(Grant No. 2013CB834606), by National Science Foundation of China (Grant Nos. 21673212, 21521001, 21473165, 21403204), by the Ministry of Science and Technology of China (Grant No. 2016YFA0400904), and by the Fundamental Research Funds for the Central Universities (Grant Nos. WK2060030018, 2030020028, 2340000074).

\bibliographystyle{apsrev}

\begin{thebibliography}{43}
\expandafter\ifx\csname natexlab\endcsname\relax\def\natexlab#1{#1}\fi
\expandafter\ifx\csname bibnamefont\endcsname\relax
  \def\bibnamefont#1{#1}\fi
\expandafter\ifx\csname bibfnamefont\endcsname\relax
  \def\bibfnamefont#1{#1}\fi
\expandafter\ifx\csname citenamefont\endcsname\relax
  \def\citenamefont#1{#1}\fi
\expandafter\ifx\csname url\endcsname\relax
  \def\url#1{\texttt{#1}}\fi
\expandafter\ifx\csname urlprefix\endcsname\relax\def\urlprefix{URL }\fi
\providecommand{\bibinfo}[2]{#2}
\providecommand{\eprint}[2][]{\url{#2}}

\bibitem[{\citenamefont{Marchetti et~al.}(2013)\citenamefont{Marchetti, Joanny,
  Ramaswamy, Liverpool, Prost, Rao, and Simha}}]{marchetti2013hydrodynamics}
\bibinfo{author}{\bibfnamefont{M.}~\bibnamefont{Marchetti}},
  \bibinfo{author}{\bibfnamefont{J.~F.} \bibnamefont{Joanny}},
  \bibinfo{author}{\bibfnamefont{S.}~\bibnamefont{Ramaswamy}},
  \bibinfo{author}{\bibfnamefont{T.~B.} \bibnamefont{Liverpool}},
  \bibinfo{author}{\bibfnamefont{J.}~\bibnamefont{Prost}},
  \bibinfo{author}{\bibfnamefont{M.}~\bibnamefont{Rao}}, \bibnamefont{and}
  \bibinfo{author}{\bibfnamefont{R.~A.} \bibnamefont{Simha}},
  \bibinfo{journal}{Reviews of Modern Physics} \textbf{\bibinfo{volume}{85}},
  \bibinfo{pages}{1143} (\bibinfo{year}{2013}).

\bibitem[{\citenamefont{Elgeti et~al.}(2015)\citenamefont{Elgeti, Winkler, and
  Gompper}}]{elgeti2015physics}
\bibinfo{author}{\bibfnamefont{J.}~\bibnamefont{Elgeti}},
  \bibinfo{author}{\bibfnamefont{R.~G.} \bibnamefont{Winkler}},
  \bibnamefont{and} \bibinfo{author}{\bibfnamefont{G.}~\bibnamefont{Gompper}},
  \bibinfo{journal}{Reports on progress in physics}
  \textbf{\bibinfo{volume}{78}}, \bibinfo{pages}{056601}
  (\bibinfo{year}{2015}).

\bibitem[{\citenamefont{Wioland et~al.}(2016)\citenamefont{Wioland, Woodhouse,
  Dunkel, and Goldstein}}]{wioland2016ferromagnetic}
\bibinfo{author}{\bibfnamefont{H.}~\bibnamefont{Wioland}},
  \bibinfo{author}{\bibfnamefont{F.~G.} \bibnamefont{Woodhouse}},
  \bibinfo{author}{\bibfnamefont{J.}~\bibnamefont{Dunkel}}, \bibnamefont{and}
  \bibinfo{author}{\bibfnamefont{R.~E.} \bibnamefont{Goldstein}},
  \bibinfo{journal}{Nature Physics} \textbf{\bibinfo{volume}{12}},
  \bibinfo{pages}{341} (\bibinfo{year}{2016}).

\bibitem[{\citenamefont{Lushi et~al.}(2014)\citenamefont{Lushi, Wioland, and
  Goldstein}}]{lushi2014fluid}
\bibinfo{author}{\bibfnamefont{E.}~\bibnamefont{Lushi}},
  \bibinfo{author}{\bibfnamefont{H.}~\bibnamefont{Wioland}}, \bibnamefont{and}
  \bibinfo{author}{\bibfnamefont{R.~E.} \bibnamefont{Goldstein}},
  \bibinfo{journal}{Proc. Natl. Acad. Sci. USA} \textbf{\bibinfo{volume}{111}},
  \bibinfo{pages}{9733} (\bibinfo{year}{2014}).

\bibitem[{\citenamefont{Goto and Tanaka}(2015)}]{goto2015purely}
\bibinfo{author}{\bibfnamefont{Y.}~\bibnamefont{Goto}} \bibnamefont{and}
  \bibinfo{author}{\bibfnamefont{H.}~\bibnamefont{Tanaka}},
  \bibinfo{journal}{Nature communications} \textbf{\bibinfo{volume}{6}},
  \bibinfo{pages}{5994} (\bibinfo{year}{2015}).

\bibitem[{\citenamefont{Szabo et~al.}(2007)}]{szabo2006phase}
\bibinfo{author}{\bibfnamefont{B.}~\bibnamefont{Szabo}},
\bibinfo{author}{\bibfnamefont{G.~J}\bibnamefont{Sz{\"o}ll{\"o}si}},
\bibinfo{author}{\bibfnamefont{B.}~\bibnamefont{G{\"o}nci}},
\bibinfo{author}{\bibfnamefont{Zs.}~\bibnamefont{Jur{\'a}nyi}},
\bibinfo{author}{\bibfnamefont{D.}~\bibnamefont{Selmeczi}},
  \bibnamefont{and}
  \bibinfo{author}{\bibfnamefont{T.}~\bibnamefont{Vicsek}},
  \bibinfo{journal}{Phys. Rev. E} \textbf{\bibinfo{volume}{74}},
  \bibinfo{pages}{061908} (\bibinfo{year}{2006}).

\bibitem[{\citenamefont{Ananthakrishnan and
  Ehrlicher}(2007)}]{ananthakrishnan2007forces}
\bibinfo{author}{\bibfnamefont{R.}~\bibnamefont{Ananthakrishnan}}
  \bibnamefont{and}
  \bibinfo{author}{\bibfnamefont{A.}~\bibnamefont{Ehrlicher}},
  \bibinfo{journal}{Int J Biol Sci} \textbf{\bibinfo{volume}{3}},
  \bibinfo{pages}{303} (\bibinfo{year}{2007}).


\bibitem[{\citenamefont{Weber et~al.}(2012)\citenamefont{Weber, Bjerke, and
  DeSimone}}]{weber2012mechanoresponsive}
\bibinfo{author}{\bibfnamefont{G.~F.} \bibnamefont{Weber}},
  \bibinfo{author}{\bibfnamefont{M.~A.} \bibnamefont{Bjerke}},
  \bibnamefont{and} \bibinfo{author}{\bibfnamefont{D.~W.}
  \bibnamefont{DeSimone}}, \bibinfo{journal}{Developmental cell}
  \textbf{\bibinfo{volume}{22}}, \bibinfo{pages}{104} (\bibinfo{year}{2012}).

\bibitem[{\citenamefont{Sato et~al.}(2015)\citenamefont{Sato, Hiraiwa, and
  Shibata}}]{sato2015cell}
\bibinfo{author}{\bibfnamefont{K.}~\bibnamefont{Sato}},
  \bibinfo{author}{\bibfnamefont{T.}~\bibnamefont{Hiraiwa}}, \bibnamefont{and}
  \bibinfo{author}{\bibfnamefont{T.}~\bibnamefont{Shibata}},
  \bibinfo{journal}{Physical review letters} \textbf{\bibinfo{volume}{115}},
  \bibinfo{pages}{188102} (\bibinfo{year}{2015}).

\bibitem[{\citenamefont{Tambe et~al.}(2011)\citenamefont{Tambe, Hardin,
  Angelini, Rajendran, Park, Serr-Picamal, Zhou, Zaman, Butler, Weitz
  et~al.}}]{tambe2011collective}
\bibinfo{author}{\bibfnamefont{D.~T.} \bibnamefont{Tambe}},
  \bibinfo{author}{\bibfnamefont{C.~C.} \bibnamefont{Hardin}},
  \bibinfo{author}{\bibfnamefont{T.~E.} \bibnamefont{Angelini}},
  \bibinfo{author}{\bibfnamefont{K.}~\bibnamefont{Rajendran}},
  \bibinfo{author}{\bibfnamefont{C.~Y.} \bibnamefont{Park}},
  \bibinfo{author}{\bibfnamefont{X.}~\bibnamefont{Serr-Picamal}},
  \bibinfo{author}{\bibfnamefont{E.~H.} \bibnamefont{Zhou}},
  \bibinfo{author}{\bibfnamefont{M.~H.} \bibnamefont{Zaman}},
  \bibinfo{author}{\bibfnamefont{J.~P.} \bibnamefont{Butler}},
  \bibinfo{author}{\bibfnamefont{D.~A.} \bibnamefont{Weitz}},
  \bibnamefont{et~al.}, \bibinfo{journal}{Nature Materials}
  \textbf{\bibinfo{volume}{10}}, \bibinfo{pages}{469} (\bibinfo{year}{2011}).

\bibitem[{\citenamefont{Levental et~al.}(2009)\citenamefont{Levental, Yu, Kass,
  Lakins, Egeblad, Erler, Fong, Csiszar, Giaccia, Weninger
  et~al.}}]{levental2009matrix}
\bibinfo{author}{\bibfnamefont{K.~R.} \bibnamefont{Levental}},
  \bibinfo{author}{\bibfnamefont{H.}~\bibnamefont{Yu}},
  \bibinfo{author}{\bibfnamefont{L.}~\bibnamefont{Kass}},
  \bibinfo{author}{\bibfnamefont{J.~N.} \bibnamefont{Lakins}},
  \bibinfo{author}{\bibfnamefont{M.}~\bibnamefont{Egeblad}},
  \bibinfo{author}{\bibfnamefont{J.~T.} \bibnamefont{Erler}},
  \bibinfo{author}{\bibfnamefont{S.~F.} \bibnamefont{Fong}},
  \bibinfo{author}{\bibfnamefont{K.}~\bibnamefont{Csiszar}},
  \bibinfo{author}{\bibfnamefont{A.}~\bibnamefont{Giaccia}},
  \bibinfo{author}{\bibfnamefont{W.}~\bibnamefont{Weninger}},
  \bibnamefont{et~al.}, \bibinfo{journal}{Cell} \textbf{\bibinfo{volume}{139}},
  \bibinfo{pages}{891} (\bibinfo{year}{2009}).

\bibitem[{\citenamefont{Bianco et~al.}(2007)\citenamefont{Bianco, Poukkula,
  Cliffe, Mathieu, Luque, Fulga, and R{\o}rth}}]{bianco2007two}
\bibinfo{author}{\bibfnamefont{A.}~\bibnamefont{Bianco}},
  \bibinfo{author}{\bibfnamefont{M.}~\bibnamefont{Poukkula}},
  \bibinfo{author}{\bibfnamefont{A.}~\bibnamefont{Cliffe}},
  \bibinfo{author}{\bibfnamefont{J.}~\bibnamefont{Mathieu}},
  \bibinfo{author}{\bibfnamefont{C.~M.} \bibnamefont{Luque}},
  \bibinfo{author}{\bibfnamefont{T.~A.} \bibnamefont{Fulga}}, \bibnamefont{and}
  \bibinfo{author}{\bibfnamefont{P.}~\bibnamefont{R{\o}rth}},
  \bibinfo{journal}{Nature} \textbf{\bibinfo{volume}{448}},
  \bibinfo{pages}{362} (\bibinfo{year}{2007}).

\bibitem[{\citenamefont{Friedl and Gilmour}(2009)}]{friedl2009collective}
\bibinfo{author}{\bibfnamefont{P.}~\bibnamefont{Friedl}} \bibnamefont{and}
  \bibinfo{author}{\bibfnamefont{D.}~\bibnamefont{Gilmour}},
  \bibinfo{journal}{Nature reviews Molecular cell biology}
  \textbf{\bibinfo{volume}{10}}, \bibinfo{pages}{445} (\bibinfo{year}{2009}).

\bibitem[{\citenamefont{Montell}(2008)}]{montell2008morphogenetic}
\bibinfo{author}{\bibfnamefont{D.~J.} \bibnamefont{Montell}},
  \bibinfo{journal}{Science} \textbf{\bibinfo{volume}{322}},
  \bibinfo{pages}{1502} (\bibinfo{year}{2008}).

\bibitem[{\citenamefont{Blum et~al.}(2014)\citenamefont{Blum, Feistel,
  Thumberger, and Schweickert}}]{blum2014evolution}
\bibinfo{author}{\bibfnamefont{M.}~\bibnamefont{Blum}},
  \bibinfo{author}{\bibfnamefont{K.}~\bibnamefont{Feistel}},
  \bibinfo{author}{\bibfnamefont{T.}~\bibnamefont{Thumberger}},
  \bibnamefont{and}
  \bibinfo{author}{\bibfnamefont{A.}~\bibnamefont{Schweickert}},
  \bibinfo{journal}{Development} \textbf{\bibinfo{volume}{141}},
  \bibinfo{pages}{1603} (\bibinfo{year}{2014}).

\bibitem[{\citenamefont{Coutelis et~al.}(2014)\citenamefont{Coutelis,
  Gonz{\'a}lez-Morales, G{\'e}minard, and Noselli}}]{coutelis2014diversity}
\bibinfo{author}{\bibfnamefont{J.-B.} \bibnamefont{Coutelis}},
  \bibinfo{author}{\bibfnamefont{N.}~\bibnamefont{Gonz{\'a}lez-Morales}},
  \bibinfo{author}{\bibfnamefont{C.}~\bibnamefont{G{\'e}minard}},
  \bibnamefont{and} \bibinfo{author}{\bibfnamefont{S.}~\bibnamefont{Noselli}},
  \bibinfo{journal}{EMBO reports} p. \bibinfo{pages}{e201438972}
  (\bibinfo{year}{2014}).

\bibitem[{\citenamefont{Wan et~al.}(2011)\citenamefont{Wan, Ronaldson, Park,
  Taylor, Zhang, Gimble, and Vunjak-Novakovic}}]{wan2011micropatterned}
\bibinfo{author}{\bibfnamefont{L.~Q.} \bibnamefont{Wan}},
  \bibinfo{author}{\bibfnamefont{K.}~\bibnamefont{Ronaldson}},
  \bibinfo{author}{\bibfnamefont{M.}~\bibnamefont{Park}},
  \bibinfo{author}{\bibfnamefont{G.}~\bibnamefont{Taylor}},
  \bibinfo{author}{\bibfnamefont{Y.}~\bibnamefont{Zhang}},
  \bibinfo{author}{\bibfnamefont{J.~M.} \bibnamefont{Gimble}},
  \bibnamefont{and}
  \bibinfo{author}{\bibfnamefont{G.}~\bibnamefont{Vunjak-Novakovic}},
  \bibinfo{journal}{Proceedings of the National Academy of Sciences}
  \textbf{\bibinfo{volume}{108}}, \bibinfo{pages}{12295}
  (\bibinfo{year}{2011}).

\bibitem[{\citenamefont{Yamanaka and Kondo}(2015)}]{yamanaka2015rotating}
\bibinfo{author}{\bibfnamefont{H.}~\bibnamefont{Yamanaka}} \bibnamefont{and}
  \bibinfo{author}{\bibfnamefont{S.}~\bibnamefont{Kondo}},
  \bibinfo{journal}{Genes to Cells} \textbf{\bibinfo{volume}{20}},
  \bibinfo{pages}{29} (\bibinfo{year}{2015}).

\bibitem[{\citenamefont{Tamada et~al.}(2010)\citenamefont{Tamada, Kawase,
  Murakami, and Kamiguchi}}]{tamada2010autonomous}
\bibinfo{author}{\bibfnamefont{A.}~\bibnamefont{Tamada}},
  \bibinfo{author}{\bibfnamefont{S.}~\bibnamefont{Kawase}},
  \bibinfo{author}{\bibfnamefont{F.}~\bibnamefont{Murakami}}, \bibnamefont{and}
  \bibinfo{author}{\bibfnamefont{H.}~\bibnamefont{Kamiguchi}},
  \bibinfo{journal}{The Journal of cell biology}
  \textbf{\bibinfo{volume}{188}}, \bibinfo{pages}{429} (\bibinfo{year}{2010}).

\bibitem[{\citenamefont{Xu et~al.}(2007)\citenamefont{Xu, Van~Keymeulen,
  Wakida, Carlton, Berns, and Bourne}}]{xu2007polarity}
\bibinfo{author}{\bibfnamefont{J.}~\bibnamefont{Xu}},
  \bibinfo{author}{\bibfnamefont{A.}~\bibnamefont{Van~Keymeulen}},
  \bibinfo{author}{\bibfnamefont{N.~M.} \bibnamefont{Wakida}},
  \bibinfo{author}{\bibfnamefont{P.}~\bibnamefont{Carlton}},
  \bibinfo{author}{\bibfnamefont{M.~W.} \bibnamefont{Berns}}, \bibnamefont{and}
  \bibinfo{author}{\bibfnamefont{H.~R.} \bibnamefont{Bourne}},
  \bibinfo{journal}{Proceedings of the National Academy of Sciences}
  \textbf{\bibinfo{volume}{104}}, \bibinfo{pages}{9296} (\bibinfo{year}{2007}).

\bibitem[{\citenamefont{Guerrero-Mart{\'\i}nez
  et~al.}(2011)\citenamefont{Guerrero-Mart{\'\i}nez, Alonso-G{\'o}mez,
  Augui{\'e}, Cid, and Liz-Marz{\'a}n}}]{guerrero2011individual}
\bibinfo{author}{\bibfnamefont{A.}~\bibnamefont{Guerrero-Mart{\'\i}nez}},
  \bibinfo{author}{\bibfnamefont{J.~L.} \bibnamefont{Alonso-G{\'o}mez}},
  \bibinfo{author}{\bibfnamefont{B.}~\bibnamefont{Augui{\'e}}},
  \bibinfo{author}{\bibfnamefont{M.~M.} \bibnamefont{Cid}}, \bibnamefont{and}
  \bibinfo{author}{\bibfnamefont{L.~M.} \bibnamefont{Liz-Marz{\'a}n}},
  \bibinfo{journal}{Nano Today} \textbf{\bibinfo{volume}{6}},
  \bibinfo{pages}{381} (\bibinfo{year}{2011}).

\bibitem[{\citenamefont{Ma et~al.}(2015)\citenamefont{Ma, Wang, Wu, and
  Wu}}]{ma2015electric}
\bibinfo{author}{\bibfnamefont{F.}~\bibnamefont{Ma}},
  \bibinfo{author}{\bibfnamefont{S.}~\bibnamefont{Wang}},
  \bibinfo{author}{\bibfnamefont{D.~T.} \bibnamefont{Wu}}, \bibnamefont{and}
  \bibinfo{author}{\bibfnamefont{N.}~\bibnamefont{Wu}},
  \bibinfo{journal}{Proceedings of the National Academy of Sciences}
  \textbf{\bibinfo{volume}{112}}, \bibinfo{pages}{6307} (\bibinfo{year}{2015}).

\bibitem[{\citenamefont{Sumino et~al.}(2012)\citenamefont{Sumino, Nagai,
  Shitaka, Tanaka, Yoshikawa, Chat{\'e}, and Oiwa}}]{sumino2012large}
\bibinfo{author}{\bibfnamefont{Y.}~\bibnamefont{Sumino}},
  \bibinfo{author}{\bibfnamefont{K.~H.} \bibnamefont{Nagai}},
  \bibinfo{author}{\bibfnamefont{Y.}~\bibnamefont{Shitaka}},
  \bibinfo{author}{\bibfnamefont{D.}~\bibnamefont{Tanaka}},
  \bibinfo{author}{\bibfnamefont{K.}~\bibnamefont{Yoshikawa}},
  \bibinfo{author}{\bibfnamefont{H.}~\bibnamefont{Chat{\'e}}},
  \bibnamefont{and} \bibinfo{author}{\bibfnamefont{K.}~\bibnamefont{Oiwa}},
  \bibinfo{journal}{Nature} \textbf{\bibinfo{volume}{483}},
  \bibinfo{pages}{448} (\bibinfo{year}{2012}).


 \bibitem[{\citenamefont{Woodhouse et~al.}(2007)\citenamefont{Woodhouse, and Goldstein}}]{woodhouse2012spontaneous}
\bibinfo{author}{\bibfnamefont{F.~G.}~\bibnamefont{Woodhouse}},
  \bibnamefont{and}
  \bibinfo{author}{\bibfnamefont{R.~E.}~\bibnamefont{Goldstein}},
  \bibinfo{journal}{Physical review letters} \textbf{\bibinfo{volume}{109}},
  \bibinfo{pages}{168105} (\bibinfo{year}{2012}).

\bibitem[{\citenamefont{Schaller et~al.}(2010)\citenamefont{Schaller, Weber,
  Semmrich, Frey, and Bausch}}]{schaller2010polar}
\bibinfo{author}{\bibfnamefont{V.}~\bibnamefont{Schaller}},
  \bibinfo{author}{\bibfnamefont{C.}~\bibnamefont{Weber}},
  \bibinfo{author}{\bibfnamefont{C.}~\bibnamefont{Semmrich}},
  \bibinfo{author}{\bibfnamefont{E.}~\bibnamefont{Frey}}, \bibnamefont{and}
  \bibinfo{author}{\bibfnamefont{A.~R.} \bibnamefont{Bausch}},
  \bibinfo{journal}{Nature} \textbf{\bibinfo{volume}{467}}, \bibinfo{pages}{73}
  (\bibinfo{year}{2010}).

\bibitem[{\citenamefont{Riedel et~al.}(2005)\citenamefont{Riedel, Kruse, and
  Howard}}]{riedel2005self}
\bibinfo{author}{\bibfnamefont{I.~H.} \bibnamefont{Riedel}},
  \bibinfo{author}{\bibfnamefont{K.}~\bibnamefont{Kruse}}, \bibnamefont{and}
  \bibinfo{author}{\bibfnamefont{J.}~\bibnamefont{Howard}},
  \bibinfo{journal}{Science} \textbf{\bibinfo{volume}{309}},
  \bibinfo{pages}{300} (\bibinfo{year}{2005}).

\bibitem[{\citenamefont{Kagami and Kamiya}(1992)}]{kagami1992translocation}
\bibinfo{author}{\bibfnamefont{O.}~\bibnamefont{Kagami}} \bibnamefont{and}
  \bibinfo{author}{\bibfnamefont{R.}~\bibnamefont{Kamiya}},
  \bibinfo{journal}{Journal of Cell Science} \textbf{\bibinfo{volume}{103}},
  \bibinfo{pages}{653} (\bibinfo{year}{1992}).

\bibitem[{\citenamefont{Jiang and Hou}(2014{\natexlab{a}})}]{jiang2014motion}
\bibinfo{author}{\bibfnamefont{H.}~\bibnamefont{Jiang}} \bibnamefont{and}
  \bibinfo{author}{\bibfnamefont{Z.}~\bibnamefont{Hou}}, \bibinfo{journal}{Soft
  Matter} \textbf{\bibinfo{volume}{10}}, \bibinfo{pages}{1012}
  (\bibinfo{year}{2014}{\natexlab{a}}).

\bibitem[{\citenamefont{Jiang and
  Hou}(2014{\natexlab{b}})}]{jiang2014hydrodynamic}
\bibinfo{author}{\bibfnamefont{H.}~\bibnamefont{Jiang}} \bibnamefont{and}
  \bibinfo{author}{\bibfnamefont{Z.}~\bibnamefont{Hou}}, \bibinfo{journal}{Soft
  matter} \textbf{\bibinfo{volume}{10}}, \bibinfo{pages}{9248}
  (\bibinfo{year}{2014}{\natexlab{b}}).

\bibitem[{\citenamefont{Dhar et~al.}(2006)\citenamefont{Dhar, Fischer, Wang,
  Mallouk, Paxton, and Sen}}]{dhar2006autonomously}
\bibinfo{author}{\bibfnamefont{P.}~\bibnamefont{Dhar}},
  \bibinfo{author}{\bibfnamefont{T.~M.} \bibnamefont{Fischer}},
  \bibinfo{author}{\bibfnamefont{Y.}~\bibnamefont{Wang}},
  \bibinfo{author}{\bibfnamefont{T.}~\bibnamefont{Mallouk}},
  \bibinfo{author}{\bibfnamefont{W.}~\bibnamefont{Paxton}}, \bibnamefont{and}
  \bibinfo{author}{\bibfnamefont{A.}~\bibnamefont{Sen}}, \bibinfo{journal}{Nano
  letters} \textbf{\bibinfo{volume}{6}}, \bibinfo{pages}{66}
  (\bibinfo{year}{2006}).

\bibitem[{\citenamefont{Qin et~al.}(2007)\citenamefont{Qin, Banholzer, Xu,
  Huang, and Mirkin}}]{qin2007rational}
\bibinfo{author}{\bibfnamefont{L.}~\bibnamefont{Qin}},
  \bibinfo{author}{\bibfnamefont{M.~J.} \bibnamefont{Banholzer}},
  \bibinfo{author}{\bibfnamefont{X.}~\bibnamefont{Xu}},
  \bibinfo{author}{\bibfnamefont{L.}~\bibnamefont{Huang}}, \bibnamefont{and}
  \bibinfo{author}{\bibfnamefont{C.~A.} \bibnamefont{Mirkin}},
  \bibinfo{journal}{Journal of the American Chemical Society}
  \textbf{\bibinfo{volume}{129}}, \bibinfo{pages}{14870}
  (\bibinfo{year}{2007}).

\bibitem[{\citenamefont{Gibbs et~al.}(2011)\citenamefont{Gibbs, Kothari,
  Saintillan, and Zhao}}]{gibbs2011geometrically}
\bibinfo{author}{\bibfnamefont{J.}~\bibnamefont{Gibbs}},
  \bibinfo{author}{\bibfnamefont{S.}~\bibnamefont{Kothari}},
  \bibinfo{author}{\bibfnamefont{D.}~\bibnamefont{Saintillan}},
  \bibnamefont{and} \bibinfo{author}{\bibfnamefont{Y.-P.} \bibnamefont{Zhao}},
  \bibinfo{journal}{Nano letters} \textbf{\bibinfo{volume}{11}},
  \bibinfo{pages}{2543} (\bibinfo{year}{2011}).

\bibitem[{\citenamefont{Kostur et~al.}(2006)\citenamefont{Kostur, Schindler,
  Talkner, and H{\"a}nggi}}]{kostur2006chiral}
\bibinfo{author}{\bibfnamefont{M.}~\bibnamefont{Kostur}},
  \bibinfo{author}{\bibfnamefont{M.}~\bibnamefont{Schindler}},
  \bibinfo{author}{\bibfnamefont{P.}~\bibnamefont{Talkner}}, \bibnamefont{and}
  \bibinfo{author}{\bibfnamefont{P.}~\bibnamefont{H{\"a}nggi}},
  \bibinfo{journal}{Physical review letters} \textbf{\bibinfo{volume}{96}},
  \bibinfo{pages}{014502} (\bibinfo{year}{2006}).

\bibitem[{\citenamefont{Speer et~al.}(2010)\citenamefont{Speer, Eichhorn, and
  Reimann}}]{speer2010exploiting}
\bibinfo{author}{\bibfnamefont{D.}~\bibnamefont{Speer}},
  \bibinfo{author}{\bibfnamefont{R.}~\bibnamefont{Eichhorn}}, \bibnamefont{and}
  \bibinfo{author}{\bibfnamefont{P.}~\bibnamefont{Reimann}},
  \bibinfo{journal}{Physical review letters} \textbf{\bibinfo{volume}{105}},
  \bibinfo{pages}{090602} (\bibinfo{year}{2010}).

\bibitem[{\citenamefont{Bogunovic et~al.}(2012)\citenamefont{Bogunovic,
  Fliedner, Eichhorn, Wegener, Regtmeier, Anselmetti, and
  Reimann}}]{bogunovic2012chiral}
\bibinfo{author}{\bibfnamefont{L.}~\bibnamefont{Bogunovic}},
  \bibinfo{author}{\bibfnamefont{M.}~\bibnamefont{Fliedner}},
  \bibinfo{author}{\bibfnamefont{R.}~\bibnamefont{Eichhorn}},
  \bibinfo{author}{\bibfnamefont{S.}~\bibnamefont{Wegener}},
  \bibinfo{author}{\bibfnamefont{J.}~\bibnamefont{Regtmeier}},
  \bibinfo{author}{\bibfnamefont{D.}~\bibnamefont{Anselmetti}},
  \bibnamefont{and} \bibinfo{author}{\bibfnamefont{P.}~\bibnamefont{Reimann}},
  \bibinfo{journal}{Physical review letters} \textbf{\bibinfo{volume}{109}},
  \bibinfo{pages}{100603} (\bibinfo{year}{2012}).

\bibitem[{\citenamefont{Nourhani et~al.}(2015)\citenamefont{Nourhani, Crespi,
  and Lammert}}]{nourhani2015guiding}
\bibinfo{author}{\bibfnamefont{A.}~\bibnamefont{Nourhani}},
  \bibinfo{author}{\bibfnamefont{V.~H.} \bibnamefont{Crespi}},
  \bibnamefont{and} \bibinfo{author}{\bibfnamefont{P.~E.}
  \bibnamefont{Lammert}}, \bibinfo{journal}{Physical Review Letters}
  \textbf{\bibinfo{volume}{115}}, \bibinfo{pages}{118101}
  (\bibinfo{year}{2015}).

\bibitem[{\citenamefont{Breier et~al.}(2016)\citenamefont{Breier, Selinger, Ciccottil, Herminghaus,
  and Mazza}}]{Breier2016spontaneous}
\bibinfo{author}{\bibfnamefont{R.~E.}~\bibnamefont{Breier}},
  \bibinfo{author}{\bibfnamefont{R.~L.~B.}~\bibnamefont{Selinger}},
  \bibinfo{author}{\bibfnamefont{G.}~\bibnamefont{Ciccotti}},
  \bibinfo{author}{\bibfnamefont{S.}~\bibnamefont{Herminghaus}},
  \bibnamefont{and} \bibinfo{author}{\bibfnamefont{M.~G.}
  \bibnamefont{Mazza}}, \bibinfo{journal}{Physical Review E}
  \textbf{\bibinfo{volume}{93}}, \bibinfo{pages}{022410}
  (\bibinfo{year}{2016}).

\bibitem[{\citenamefont{Ahlrichs and D{\"u}nweg}(1998)}]{ahlrichs1998lattice}
\bibinfo{author}{\bibfnamefont{P.}~\bibnamefont{Ahlrichs}} \bibnamefont{and}
  \bibinfo{author}{\bibfnamefont{B.}~\bibnamefont{D{\"u}nweg}},
  \bibinfo{journal}{International Journal of Modern Physics C}
  \textbf{\bibinfo{volume}{9}}, \bibinfo{pages}{1429} (\bibinfo{year}{1998}).

\bibitem[{\citenamefont{Mones et~al.}(2015)\citenamefont{Mones, Czir{\'o}k, and
  Vicsek}}]{mones2015anomalous}
\bibinfo{author}{\bibfnamefont{E.}~\bibnamefont{Mones}},
  \bibinfo{author}{\bibfnamefont{A.}~\bibnamefont{Czir{\'o}k}},
  \bibnamefont{and} \bibinfo{author}{\bibfnamefont{T.}~\bibnamefont{Vicsek}},
  \bibinfo{journal}{New journal of physics} \textbf{\bibinfo{volume}{17}},
  \bibinfo{pages}{063013} (\bibinfo{year}{2015}).

\bibitem[{\citenamefont{Kozlov and Mogilner}(2007)}]{kozlov2007model}
\bibinfo{author}{\bibfnamefont{M.~M.} \bibnamefont{Kozlov}} \bibnamefont{and}
  \bibinfo{author}{\bibfnamefont{A.}~\bibnamefont{Mogilner}},
  \bibinfo{journal}{Biophysical journal} \textbf{\bibinfo{volume}{93}},
  \bibinfo{pages}{3811} (\bibinfo{year}{2007}).

\bibitem[{\citenamefont{Yam et~al.}(2007)\citenamefont{Yam, Wilson, Ji, Hebert,
  Barnhart, Dye, Wiseman, Danuser, and Theriot}}]{yam2007actin}
\bibinfo{author}{\bibfnamefont{P.~T.} \bibnamefont{Yam}},
  \bibinfo{author}{\bibfnamefont{C.~A.} \bibnamefont{Wilson}},
  \bibinfo{author}{\bibfnamefont{L.}~\bibnamefont{Ji}},
  \bibinfo{author}{\bibfnamefont{B.}~\bibnamefont{Hebert}},
  \bibinfo{author}{\bibfnamefont{E.~L.} \bibnamefont{Barnhart}},
  \bibinfo{author}{\bibfnamefont{N.~A.} \bibnamefont{Dye}},
  \bibinfo{author}{\bibfnamefont{P.~W.} \bibnamefont{Wiseman}},
  \bibinfo{author}{\bibfnamefont{G.}~\bibnamefont{Danuser}}, \bibnamefont{and}
  \bibinfo{author}{\bibfnamefont{J.~A.} \bibnamefont{Theriot}},
  \bibinfo{journal}{The Journal of cell biology}
  \textbf{\bibinfo{volume}{178}}, \bibinfo{pages}{1207} (\bibinfo{year}{2007}).

\bibitem[{\citenamefont{Dawes and
  Edelstein-Keshet}(2007)}]{dawes2007phosphoinositides}
\bibinfo{author}{\bibfnamefont{A.~T.} \bibnamefont{Dawes}} \bibnamefont{and}
  \bibinfo{author}{\bibfnamefont{L.}~\bibnamefont{Edelstein-Keshet}},
  \bibinfo{journal}{Biophysical journal} \textbf{\bibinfo{volume}{92}},
  \bibinfo{pages}{744} (\bibinfo{year}{2007}).

\bibitem[{\citenamefont{Szab{\'o} et~al.}(2010)\citenamefont{Szab{\'o},
  {\"U}nnep, M{\'e}hes, Twal, Argraves, Cao, and
  Czir{\'o}k}}]{szabo2010collective}
\bibinfo{author}{\bibfnamefont{A.}~\bibnamefont{Szab{\'o}}},
  \bibinfo{author}{\bibfnamefont{R.}~\bibnamefont{{\"U}nnep}},
  \bibinfo{author}{\bibfnamefont{E.}~\bibnamefont{M{\'e}hes}},
  \bibinfo{author}{\bibfnamefont{W.}~\bibnamefont{Twal}},
  \bibinfo{author}{\bibfnamefont{W.}~\bibnamefont{Argraves}},
  \bibinfo{author}{\bibfnamefont{Y.}~\bibnamefont{Cao}}, \bibnamefont{and}
  \bibinfo{author}{\bibfnamefont{A.}~\bibnamefont{Czir{\'o}k}},
  \bibinfo{journal}{Physical biology} \textbf{\bibinfo{volume}{7}},
  \bibinfo{pages}{046007} (\bibinfo{year}{2010}).

\bibitem[{\citenamefont{Speck et~al.}(2014)\citenamefont{Speck, Bialk{\'e},
  Menzel, and L{\"o}wen}}]{speck2014effective}
\bibinfo{author}{\bibfnamefont{T.}~\bibnamefont{Speck}},
  \bibinfo{author}{\bibfnamefont{J.}~\bibnamefont{Bialk{\'e}}},
  \bibinfo{author}{\bibfnamefont{A.~M.} \bibnamefont{Menzel}},
  \bibnamefont{and}
  \bibinfo{author}{\bibfnamefont{H.}~\bibnamefont{L{\"o}wen}},
  \bibinfo{journal}{Physical Review Letters} \textbf{\bibinfo{volume}{112}},
  \bibinfo{pages}{218304} (\bibinfo{year}{2014}).

\bibitem[{\citenamefont{Ni et~al.}(2015)\citenamefont{Ni, Stuart, and
  Bolhuis}}]{ni2015tunable}
\bibinfo{author}{\bibfnamefont{R.}~\bibnamefont{Ni}},
  \bibinfo{author}{\bibfnamefont{M.~A.~C.} \bibnamefont{Stuart}},
  \bibnamefont{and} \bibinfo{author}{\bibfnamefont{P.~G.}
  \bibnamefont{Bolhuis}}, \bibinfo{journal}{Physical review letters}
  \textbf{\bibinfo{volume}{114}}, \bibinfo{pages}{018302}
  (\bibinfo{year}{2015}).

\bibitem[{\citenamefont{Stenhammar et~al.}(2015)\citenamefont{Stenhammar,
  Wittkowski, Marenduzzo, and Cates}}]{stenhammar2015activity}
\bibinfo{author}{\bibfnamefont{J.}~\bibnamefont{Stenhammar}},
  \bibinfo{author}{\bibfnamefont{R.}~\bibnamefont{Wittkowski}},
  \bibinfo{author}{\bibfnamefont{D.}~\bibnamefont{Marenduzzo}},
  \bibnamefont{and} \bibinfo{author}{\bibfnamefont{M.~E.} \bibnamefont{Cates}},
  \bibinfo{journal}{Physical review letters} \textbf{\bibinfo{volume}{114}},
  \bibinfo{pages}{018301} (\bibinfo{year}{2015}).

\end{thebibliography}

\end{document}